\documentstyle[12pt,epsf,axodraw]{article}
\newcommand{\beq}{\begin{eqnarray}}
\newcommand{\eeq}{\end{eqnarray}}

\newcommand{\bsg}{ b \to s g}

\newcommand{\bsqq}{b \to s q \bar{q}}

\newcommand{\brbsg}{ BR( b \to s g) }

\newcommand{\brnoc}{BR( b \to no \ \ charm)}
\newcommand{\bsga}{      b \to s \gamma }
\newcommand{\brbsga}{BR( b \to s \gamma )}
\newcommand{\bsl}{  B_{SL}}
\newcommand{\nc}{ n_c}

\newcommand{\mhp}{M_{H^{+}}}

\newcommand{\tab}[1]{Table \ref{#1}}
\newcommand{\rnc}{r_{ c \hspace{-1.1truemm}/ }}

\newcommand{\real}{{\rm Re}\,}
\title{ {\bf Charm multiplicity and the branching ratios of inclusive
charmless b quark decays in the general two-Higgs-doublet models}}
\author{ Zhenjun Xiao
\thanks{E-mail: zxiao@ibm320h.phy.pku.edu.cn},
Chong Sheng Li \\
{\small Department of Physics, Peking University, Beijing,
 100871 P.R. China }\\
 Kuang-Ta Chao \\
{\small  CCAST(World Laboratory),P.O.Box 8730, Beijing 100080,
P.R.China }\\
{\small and Department of Physics, Peking University, Beijing,
100871 P.R. China}   \\ }
\date{\today}

\begin{document}
\maketitle
\begin{abstract}
In the framework of general two-Higgs-doublet models, we calculate
the branching ratios of various inclusive charmless b decays by
using the low energy effective Hamiltonian including next-to-leading
order QCD corrections, and examine the current status and the new
physics effects on the determination of the charm multiplicity $n_c$
and semileptonic branching ratio  $B_{SL}$. Within the considered
parameter space, the enhancement to the ratio $BR(b \to s g)$
due to the charged-Higgs penguins can be as large as a factor of
8 (3) in the model III (II), while the ratio $BR(b \to no \ \ charm)$
can be increased from the standard model  prediction of $2.49\%$
to $4.91\%$ ($2.99\%$) in the model III (II). Consequently, the value
of $B_{SL}$ and $n_c$ can be decreased simultaneously in the model
III. The central value of $B_{SL}$ will be lowered slightly by about
$0.003$, but the ratio $n_c$ can be reduced significantly from the
theoretical prediction of $n_c= 1.28 \pm 0.05$ in the SM to $n_c=
1.23 \pm 0.05$, $1.18 \pm 0.05$ for $m_{H^+}=200, 100$ GeV,
respectively. We find that the  predicted $n_c$ and the measured
$n_c$ now agree within roughly one standard deviation after taking
into account the effects of gluonic charged Higgs penguins in the
model III with a relatively light charged Higgs boson.
\end{abstract}


\vspace{.5cm}

\newpage

\section*{ I. Introduction}

In the forthcoming years, experiments at SLAC and KEK B-factories,
HERA-B and other high energy colliders  will measure various
branching ratios and CP-violating asymmetries of B decays
\cite{slac504,ellis99}. The expected large
number of B decay events ( say $10^8 - 10^9$) may allow us to explore
the physics of CP violation, to determine the flavor parameters of
the electroweak theory, and to probe for signals or evidences of
new physics beyond the Standard Model (SM) [1 - 6].

Among various B meson decay modes, the  decay $b \to s \gamma$ and
$b \to s g$ have been, for example, the hot subject of many
investigations \cite{ciuchini94}, since these decay
modes may be affected by loop contributions from various new physics
models. Great progress in both the theoretical calculation
\cite{kagan99} and
the experimental measurement \cite{cleo99} enable one to
constrain the new physics models, such as the two-Higgs-doublet
model (2HDM) \cite{2hdm}, the minimal supersymmetric standard
model \cite{misiak97} and the Technicolor models \cite{xiao96}.

For many years, it appeared that the SM prediction for the
semileptonic branching ratio $\bsl$ \cite{bigi94} is much larger
than the values measured at $\Upsilon$ resonance and $Z^0$-peak
\cite{cleo96,pdg98}. More recently, the
theoretical predictions have been refined by including full $O(\alpha_s)$
QCD corrections \cite{bslsm,neubert97}. These progress, consequently,
have lowered the predicted $\bsl$ and now adequately reproduce
the experimental results \cite{pdg98}. However, the measurements
of $\bsl$ obtained at the $\Upsilon(4S)$ and $Z^0$ resonance are
still disagree slightly \cite{yamamoto99}.
Besides the $\bsl$ problem, there is another so-called "missing
charm puzzle" \cite{pdg98,kagan98}: the charm multiplicity $\nc$
measured at CLEO and LEP \cite{yamamoto99,cleo97}( especially at
CLEO, the $\Upsilon$ resonance )  is smaller than the theoretical
prediction. Among various possible
explanations for the missing charm/$\bsl$ problem, the most
intriguing one would be an enhanced $B \to X_{no charm}$
rate due to new physics beyond the SM \cite{kagan98}. An enhanced
$b \to sg$ can decrease the values of both $\nc$ and the $\bsl$
simultaneously \cite{kagan98}. The large branching ratio
$BR(B \to \eta' X_s)$ reported recently by CLEO \cite{cleo98}
provided a new hint for enhanced $b \to sg$. Besides those
explanations based on the SM \cite{etapsm}, new physics interpretation
for this large ratio is also plausible \cite{tseng98}.

In a previous paper \cite{xiao992}, we calculated, from the first
principle, the new contributions to inclusive charmless b quark decays $\bsg,
\bsqq$ from the gluonic charged-Higgs penguin diagrams in the
so-called Model III: the two-Higgs-doublet model
with flavor changing  couplings \cite{hou92,atwood97}. In the
considered parameter space, we found that the branching ratio
$BR(b \to sg)$ ($q^2 =0$) can be increased by roughly an order of
magnitude, which is much larger than
that in the ordinary 2HDM's \cite{hou94}. In \cite{xiao992},
however,  we used the language of form factors $F_1$ and $F_2$ and
took into account the QCD corrections partially by using the
$\alpha_s(m_b)$ directly to calculate the branching ratios.

In this paper, in the framework of general 2HDM's, we will calculate
the branching ratios of various inclusive charmless b decays by using
the low energy effective Hamiltonian including
next-to-leading order (NLO) QCD corrections \cite{lenz97}, and
investigate the new physics effects on the theoretical predictions
for both $\bsl$ and $\nc$.

This paper is organized as follows. In Sec.II, we describe the basic
structures  of the model III, extract out the Wilson coefficients,
draw the constraint on parameter space of the model III from
currently available data. In Sec.III, we calculate the branching
ratios $\brbsg$ and  $BR(b \to q' q\bar{q} )$ for $q'\in {d,s}$
and $q\in {u,d,s}$ in the model III and II  with the inclusion of
NLO QCD corrections. In Sec.IV, we examine the current status and
new physics effects on the determination of $\bsl$ and $\nc$.
The conclusions and discussions are included in the final section.

\section*{II. The general 2HDM's and experimental constraint}

The simplest extension of the SM is the so-called two-Higgs-doublet
models\cite{2hdm}. In such models, the tree level flavor changing
neutral currents(FCNC's)are absent if one introduces an {\it ad
hoc} discrete symmetry to constrain the 2HDM scalar potential and
Yukawa Lagrangian. Lets consider a Yukawa Lagrangian
of the form\cite{atwood97}
\beq
{\cal L}_Y &=&
\eta^U_{ij}\bar{Q}_{i,L} \tilde{\phi_1}U_{j,R} +
\eta^D_{ij}\bar{Q}_{i,L} \phi_1 D_{j,R}
+\xi^U_{ij}\bar{Q}_{i,L} \tilde{\phi_2}U_{j,R}
+\xi^D_{ij}\bar{Q}_{i,L} \phi_2 D_{j,R}+ h.c., \label{leff}
\eeq
where $\phi_{i}$ ($i=1,2$) are the two Higgs doublets of a
two-Higgs-doublet model, $\tilde{\phi}_{1,2}=
i\tau_2 \phi^*_{1,2}$, $Q_{i,L}$ ($U_{j,R}$) with $i=(1,2,3)$ are the
left-handed isodoublet quarks (right-handed   up-type quarks),
$D_{j,R}$  are the right-handed  isosinglet  down-type quarks,
while $\eta^{U,D}_{i,j}$  and $\xi^{U,D}_{i,j}$ ($i,j=1,2,3$ are
family index ) are generally the nondiagonal matrices of the  Yukawa
coupling. By imposing the discrete symmetry
\beq
\phi_1 \to - \phi_1, \phi_2 \to \phi_2,
D_i \to - D_i, U_i \to  \mp U_i
\eeq
one obtains the so called Model I and Model II. In Model I the third
and fourth term in eq.(\ref{leff}) will be dropped by the discrete
symmetry, therefore, both the up- and down-type quarks get mass from
Yukawa couplings to the same Higgs doublet $\phi_1$, while the
$\phi_2$ has no Yukawa couplings to the quarks. For Model II, on
the other hand, the first and fourth term in
Eq.(\ref{leff}) will be dropped by imposing the discrete symmetry.
Model II has, consequently the up- and down-type quarks getting mass
from Yukawa couplings to two different scalar doublets $\phi_1$ and
$\phi_2$.

During past years, the models I and II have been studied extensively in
literature and tested experimentally, and the model II has been very
popular since it is the building block of the minimal supersymmetric
standard model. In this paper, we focus on  the third
type of 2HDM \cite{hou92}, usually known as the model III
\cite{hou92,atwood97}. In the model III, no discrete symmetry is
imposed and both up- and down-type quarks then may have diagonal
and/or flavor changing couplings with $\phi_1$ and $\phi_2$.
As described in \cite{atwood97}, one can choose a suitable
basis $(H^0, H^1, H^2, H^\pm)$ to express two Higgs
doublets \cite{atwood97}
\beq
\phi_1 = \frac{1}{\sqrt{2}}\left ( \begin{array}{c}
\sqrt{2} \chi^+ \\ v + H^0 + i \chi^0\\ \end{array} \right ),\ \
\phi_2=\frac{1}{\sqrt{2}}\left ( \begin{array}{c}
\sqrt{2} H^+ \\ H^1 + i H^2 \\ \end{array} \right ),
\label{phi12}
\eeq
and take their vacuum expectation values as the form
\beq
<\phi_1> &=& \left ( \begin{array}{c}
0 \\ v/\sqrt{2}\\ \end{array} \right ), \ \  <\phi_2>=0,
\label{vev}
\eeq
where $v=(\sqrt{2}G_F)^{-1/2}=246GeV$. The transformation relation between $(H^0,H^1,H^2)$
and the mass eigenstates $(\overline{H}^0, h^0, A^0)$ can be found
in \cite{atwood97}. The $H^\pm$ are the physical charged Higgs
boson, $H^0$ and $h^0$ are the physical CP-even neutral Higgs boson
and the $A^0$ is the physical CP-odd neutral Higgs boson. After the
rotation of quark fields, the Yukawa Lagrangian of quarks are of the
form \cite{atwood97},
\beq
{\cal L}_Y^{III} =
\eta^U_{ij}\bar{Q}_{i,L} \tilde{\phi_1}U_{j,R} +
\eta^D_{ij}\bar{Q}_{i,L} \phi_1 D_{j,R}
+\hat{\xi}^U_{ij}\bar{Q}_{i,L} \tilde{\phi_2}U_{j,R}
+\hat{\xi}^D_{ij}\bar{Q}_{i,L} \phi_2 D_{j,R} + H.c.,
\label{lag3}
\eeq
where $\eta^{U,D}_{ij}$ correspond to the diagonal mass matrices of
up- and down-type quarks, while the neutral and charged flavor changing
couplings will be \cite{atwood97}
\footnote{We make the same ansatz on the $\xi^{U,D}_{ij}$ couplings as
the Ref.\cite{atwood97}. For more details about the definition of
$\hat{\xi}^{U,D}$ one can see Ref.\cite{atwood97}. }
\beq
\xi^{U,D}_{ij}=\frac{\sqrt{m_im_j}}{v} \lambda_{ij}, \ \
\hat{\xi}^{U,D}_{neutral}= \xi^{U,D}, \ \
\hat{\xi}^{U}_{charged}= \xi^{U}V_{CKM}, \ \
\hat{\xi}^{D}_{charged}= V_{CKM} \xi^{D}, \label{cxiud}
\eeq
where $V_{CKM}$ is the Cabibbo-Kabayashi-Maskawa mixing matrix
\cite{ckm}, $i,j=(1,2,3)$ are the generation index. The coupling
constants $\lambda_{ij}$ are free parameters to be determined by
experiments, and they may also be complex.

In the model II and assuming $\tan{\beta} =1 $, the constraint on the mass
of charged Higgs boson due to CLEO data of $\bsga$ is $\mhp \geq 350$
( 200 ) GeV at the LO (NLO) level \cite{lo2hdm,nlo2hdm}. For the model
I, however, the limit can be much weaker due to the possible
destructive interference with the SM amplitude.

For the model III, the situation is not as clear as the model II
because there are more free parameters here. As pointed in
\cite{atwood97},  the data of $K^0-\bar{K}^0$ and $B_d^0-
\bar{B}_d^0$ mixing processes put severe constraint on the FC
couplings involving the first generation of quarks. One therefore
assume that,
\beq
\lambda_{uj}=\lambda_{dj}=0, \ \ for \ \ j=1,2,3 \label{limit1}
\eeq

Imposing the limit in Eq.(\ref{limit1}) and assuming all other $\lambda_{ij}$
parameters are of order 1, Atwood et al. \cite{atwood96} found a very
strong constraint of  $\mhp > 600 GeV$ by using the CLEO data of
$\bsga$ decay available in 1995. In Ref.\cite{aliev99},
Aliev et al. studied the
$\bsga$ decay in the model III by extending the NLO results of the
model II \cite{nlo2hdm} to the case of model III, and found some
constraints on the FC couplings.

In a recent paper \cite{chao99}, Chao et al., studied the decay $\bsga$ by
assuming that only the couplings $\lambda_{tt}$ and $\lambda_{bb}$ are
non-zero. They found that the constraint on $\mhp$ imposed by the CLEO
data of $\bsga$ can be greatly relaxed by considering the phase
effects of $\lambda_{tt}$ and $\lambda_{bb}$. The constraints by
$B^0-\overline{B^0}$ mixing, the neutron electric dipole moment(NEDM),
the $Z^0$-pole parameter $\rho$ and $R_b$ give the following
preferred scenario \cite{chao99}:
\beq
|\lambda_{tt}| \leq  0.3, \ \ |\lambda_{bb}| \approx 50, \ \
 M_{A^0} \approx M_{h^0} =80 - 120 GeV; \ \
 80 GeV \leq \mhp \leq  200 GeV \label{limit2}
\eeq

In the following sections, we will calculate the new physics
contributions to the inclusive charmless decays of b quark in the
Chao-Cheung-Keung (CCK) scenario of model III \cite{chao99}. Such
model III has following advantages:

\begin{enumerate}
\item
Since we keep only the couplings $\lambda_{tt}$ and $\lambda_{bb}$ none zero,
the neutral Higgs bosons do not contribute at tree level or one-loop
level. The new contributions therefore come only from the charged Higgs
penguin diagrams with the heavy internal top quark.

\item
The new operators $O_{9,10}$ and all flipped chirality partners of
operators $O_{1, \cdots,10}$ as defined in \cite{aliev99} do not
contribute to the decay $\bsga$ and other inclusive charmless decays
under study in this paper.

\item
The free parameters in this model III are greatly reduced to $\lambda_{tt}$,
$\lambda_{bb}$ and $\mhp$.

\end{enumerate}

In order to find more details about the correlations between $\mhp$ and
couplings $\lambda_{tt,bb}$ by imposing the new CLEO data of $\bsga$,
we recalculate the decay $\bsga$ in the model III. For the sake of
simplicity, we do not consider the less interesting model I further
in this paper.

The effective Hamiltonian for $B \to X_S \gamma$ at the scale
$\mu = O(m_b)$ is given by \cite{buras97}
\beq
{\cal H}_{eff}(b \to s \gamma) = -\frac{G_F}{\sqrt{2}} V_{ts}^* V_{tb}
\left [ \sum_{i=1}^6 C_i(\mu) Q_i(\mu) + C_{7\gamma}(\mu) Q_{7\gamma}
+ C_{8G}(\mu) Q_{8G} \right ]\label{heffbsga}
\eeq
The explicit expressions of operators $Q_{1-6}, Q_{7\gamma}$ and
$Q_{8G}$,  as well as the corresponding Wilson coefficients $C_i(M_W)$
in the SM can be found for example in \cite{buras97}.

In the model III, the left-handed QED magnetic-penguin operator
$Q_{7\gamma}^L$ and the left-handed  QCD magnetic-penguin operator
$Q_{8G}^L$ may also play an important role,
\beq
Q_{7\gamma}^{L}&=& \frac{e}{8\pi^2}m_b\bar{s}_{\alpha} \sigma^{\mu \nu}
( 1 - \gamma_5) b_{\alpha} F_{\mu \nu},\label{q7lnew} \\
Q_{8G}^{L}&=& \frac{g_s}{8\pi^2}m_b\bar{s}_{\alpha} \sigma^{\mu \nu}
( 1 - \gamma_5) T^a_{\alpha \beta} b_{\beta} G^a_{\mu \nu},\label{q8lnew}
\eeq
In the SM and ordinary 2HDM's, both operators $Q_{7\gamma}^L$ and
$Q_{8G}^L$  are  absent because one usually assume that
$m_s/m_b \sim 0$. In the model III, however, these two  left-handed
operators may contribute effectively  because the Wilson
coefficients $C_{7\gamma}^L$ and $C_{8G}^L$ may be rather large
to compensate for the suppression of $m_s/m_b$.

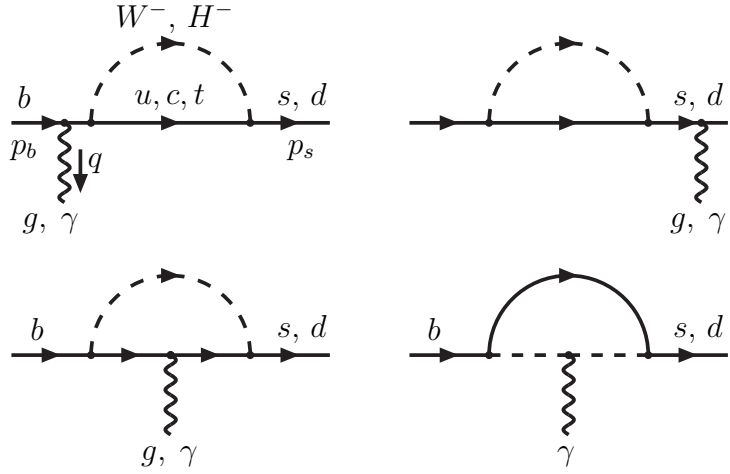
\begin{figure}[t]
\vspace{1cm}\hfill \\
\SetScale{1}
\begin{picture}(70,30)(0,0)
\SetWidth{1.4}
\ArrowLine(50,50)(80,50)  \Vertex(80,50){1.5} \Text(55,60)[]{$b$}
\ArrowLine(80,50)(140,50) \Text(110,60)[]{$u, c, t $}
\Vertex(140,50){1.5} \Text(112,90)[]{$W^-,\, H^-$}
\ArrowLine(140,50)(170,50) \Text(160,60)[]{$s,\, d$} \Text(55,40)[]{$p_b$}
\Photon(70,50)(70,20){2}{4} \Text(65,12)[]{$g,\; \gamma$} \Vertex(70,50){1.5}
\LongArrow(76,40)(76,25) \Text(82,35)[]{$q$}
\DashArrowArcn(110,50)(30,180,360){5} \Text(160,40)[]{$p_s$}
\ArrowLine(200,50)(230,50)  \Vertex(230,50){1.5}
\ArrowLine(230,50)(290,50) \Vertex(290,50){1.5} 
\ArrowLine(290,50)(320,50) \Text(310,60)[]{$s,\, d$} 
\DashArrowArcn(260,50)(30,180,360){5}     
\Photon(310,50)(310,20){2}{4}  \Text(310,12)[]{$g,\; \gamma$}
\Vertex(310,50){1.5} 
\end{picture}
\vspace{2cm} \hfill \\
\SetScale{1}
\begin{picture}(70,30)(0,0)
\SetWidth{1.4}
\ArrowLine(50,50)(80,50)  \Vertex(80,50){1.5} \Text(60,60)[]{$b$}
\ArrowLine(80,50)(110,50) \Vertex(110,50){1.5}
\ArrowLine(110,50)(140,50) \Vertex(140,50){1.5}
\ArrowLine(140,50)(170,50) \Text(160,60)[]{$s,\, d$}
\DashArrowArcn(110,50)(30,180,360){5}
\Photon(110,50)(110,20){2}{4}  \Vertex(110,50){1.5}
\Text(110,12)[]{$g,\; \gamma$}
\ArrowLine(200,50)(230,50) \Vertex(230,50){1.5} \Text(210,60)[]{$b$}
\DashLine(230,50)(260,50){5} \Vertex(260,50){1.5}
\DashLine(260,50)(290,50){5} \Vertex(290,50){1.5}
\ArrowLine(290,50)(320,50)   \Text(310,60)[]{$s,\, d$}
\ArrowArcn(260,50)(30,180,360)
\Photon(260,50)(260,20){2}{4} \Text(260,12)[]{$\gamma$}
\end{picture}
\caption{The Feynman diagrams for the decays $b \to s \gamma$  and
$b \to s g$ in the SM and 2HDM's. The internal quarks are the upper
type $u, c$ and $t$ quarks.}
\label{fig:fig1}
\end{figure}

In Ref.\cite{xiao992}, we calculated the $\bsg$ decay in the model III from
the first principle and obtained the corresponding form factors $F_1$ and
$F_2$. Following the standard procedure and using the Feynman rules
in the model III \cite{atwood97}, we evaluate the
Feynman diagrams for both $\bsga$ and $\bsg$ decay as shown in
Fig.\ref{fig:fig1}, extract out the Wilson coefficients
$C_i(M_W)$ at the energy scale $M_W$ by matching the full theory onto the
effective theory,
\beq
C_{i}(M_W)&=& 0 \ \  (i=1,3,4,5,6),\label{31mw}\\
C_{2}(M_W)&=& 1,  \label{32mw}\\
C_{7\gamma}^{L}(M_W)&=& -\frac{m_s}{18 m_b}D(y_t)|\lambda_{tt}|^2,
    \label{37lmw}\\
C_{7\gamma}^R(M_W)&=& C_{7\gamma}(M_W)^{SM}
    - \frac{1}{12}A(y_t)|\lambda_{tt}|^2
    + \frac{1}{2} B(y_t)|\lambda_{tt}\lambda_{bb}| e^{\theta},
    \label{37rmw}\\
C_{8G}^L(M_W)&=& - \frac{m_s}{12 m_b}D(y_t)|\lambda_{tt}|^2, \label{38lmw}\\
C_{8G}^R(M_W)&=& C_{8G}(M_W)^{SM}
    - \frac{1}{12}D(y_t)|\lambda_{tt}|^2
    + \frac{1}{2} E(y_t)|\lambda_{tt}\lambda_{bb}| e^{\theta},
    \label{38rmw}
\eeq
with
\beq
C_{7\gamma}(M_W)^{SM}&=& -\frac{A(x_t)}{2}, \label{c7mwsm}\\
C_{8G}(M_W)^{SM}&=& -\frac{D(x_t)}{2}, \label{c8mwsm}
\eeq
where $x_t=m_t^2/M_W^2$, $y_t=m_t^2/\mhp^2$, the phase angle
$\theta=\theta_b-\theta_t$, while $\theta_b$ ( $\theta_t$) is
the phase angle of $\lambda_{bb}$ ($\lambda_{tt}$). When compared
with the Eqs.(18,19) of Ref.\cite{chao99}, the second and third terms
in Eqs.(\ref{37rmw}) and (\ref{38rmw}) have an additional factor of
$1/2$, since  $\xi^{U,D}_{ij}$ used here has as additional factor
$1/\sqrt{2}$.  The Inami-Lim functions \cite{inami81} $(A, B, D, E)$
are of the form,
\beq
A(x)&=&\frac{7x - 5x^2 - 8x^3}{12(1-x)^3}
    + \frac{2x^2 -3x^3}{2(1-x)^4}\log[x], \label{ax}\\
D(x)&=&\frac{2x + 5x^2 - x^3}{4(1-x)^3}
    + \frac{3x^2}{2(1-x)^4}\log[x],  \label{dx}\\
B(y)&=&\frac{-3y + 5y^2}{12(1-y)^2}
    - \frac{2y -3y^2}{6(1-y)^3}\log[y], \label{by} \\
E(y)&=&\frac{-3y + y^2}{4(1-y)^2} - \frac{y}{2(1-y)^3}\log[y]
\label{ey}
\eeq
The Wilson coefficients given in Eqs.(\ref{31mw}-\ref{38rmw}) contained
the contributions from both the $W^\pm$-penguin and $H^\pm$-penguin
diagrams.

It is easy to see that both $C_{7\gamma}^L(M_W)$ and $C_{8G}^L(M_W)$ in
Eqs.(\ref{37lmw}) and (\ref{38lmw}) will be doubly suppressed by the ratio
$m_s/m_b$ and  $|\lambda_{tt}|^2$ when $|\lambda_{tt}|$ is small as
preferred by the data of NEDM \cite{chao99}. For typical values of
relevant parameters, say
$|\lambda_{tt}|=0.3$, $|\lambda_{bb}|=40$, $\theta=0^0$ and
$\mhp=200$ GeV, One finds numerically that $C_{7\gamma}^L(M_W)\approx
C_{8G}^L(M_W)) \approx 10^{-5}$, while $C_{7\gamma}^R(M_W) \approx
C_{8G}^R(M_W) \approx 0.8$. Consequently, the left-handed Wilson
coefficients are much smaller than their right-handed counterparts
and therefore  will be neglected in the following calculations.

At the lower energy  scale $\mu =O(m_b)$, the  Wilson coefficients
$C_i(\mu)$ for the decay $\bsga$ at the leading order are of the
form
\beq
C_j(\mu) &=& \sum_{i=1}^6 k_{ji} \eta^{a_i}\ \
    (j=1,\cdots, 6),\label{cimu} \\
C_{7\gamma}(\mu)^{SM}&=& \eta^{\frac{16}{23}} C_{7\gamma}(M_W)^{SM}
  + \frac{8}{3} \left ( \eta^{\frac{14}{23}} - \eta^{\frac{16}{23}}\right )
     C_{8G}(M_W)^{SM}+ \sum_{i=1}^8 h_i \eta^{a_i}, \label{c7sm} \\
C_{7\gamma}(\mu)^{III}&=& \eta^{\frac{16}{23}} C_{7\gamma}^R(M_W)
  + \frac{8}{3} \left ( \eta^{\frac{14}{23}} - \eta^{\frac{16}{23}}\right )
     C_{8G}^R(M_W)+ \sum_{i=1}^8 h_i \eta^{a_i}, \label{c7r3}
\eeq
where $\eta=\alpha_s(M_W)/\alpha_s(\mu)$, and the scheme-independent
numbers $a_i$, $k_{ji}$ and $h_i$ can be found in \cite{buras97}.

Using the effective Hamiltonian,  the branching ratio of $\bsga$ at
the leading order can be written as,
\beq
\brbsga^{(III)} &=&\frac{|V^*_{ts}V_{tb}|^2}{|V_{cb}|^2}
\frac{6 \alpha_{em}}{\pi f(z)} |C_{7\gamma}(\mu)^{III}|^2
BR(b \to ce\bar{\nu}), \label{brbsga3}
\eeq
where $\mu =O(m_b)$, $BR(b \to ce \bar{\nu})=(10.7 \pm 0.4)\%$ is
the measured semileptonic branching ratio of b decay, and $f(z)$
is the  phase space factor,
\beq
f(z)&=&1-8z^2 + 8 z^6 -z^8 -24z^4 \log[z], \label{fz}
\eeq
where $z=m_c^{pole}/m_b^{pole}$. It is straightforward to write
down the branching ratios $\brbsga$ for the SM and model II.

In the numerical calculations, the following input parameters
\cite{pdg98,buras98} will  be used implicitly:
\beq
M_W&=&80.41GeV,\ \  M_Z=91.187GeV,\ \  \alpha_{em}=1/137,\nonumber\\
\alpha_s(M_Z)&=& 0.118, \ \ G_F=1.16639\times 10^{-5} (GeV)^{-2}, \nonumber\\
m_s&=&0.13 GeV,\ \ m_c=1.4 GeV,\ \  m_b =4.8 GeV, \nonumber\\
m_t&=&\overline{m_t}(m_t)=168 GeV,  \ \
\Lambda^{(5)}_{\overline{MS}}=0.225, \nonumber\\
A&=&0.84,\ \ \lambda=0.22, \ \ \rho=0.20, \ \ \eta=0.34,
\label{sip}
\eeq
where $A, \lambda, \rho$ and $\eta$ are the Wolfenstein parameters
of the CKM mixing matrix. $\overline{m_t}(m_t)$ here refers to the
running current top quark mass normalized at $\mu=m_t$ and is
obtained from the pole mass $m_t^{pole}=176$ GeV. For the running of
$\alpha_s$, the two-loop formulae \cite{buras97} will be used.

Fig.2 shows the branching ratios $\brbsga$ in the SM and models II and
III, assuming $\lambda_{tt}=0.3$,$\lambda_{bb}=35$,
$\theta=0^0, 30^0$, $\tan{\beta}=1$. The horizontal band between
two dotted lines corresponds to the CLEO data \cite{cleo99}:
$2\times 10^{-4} \leq \brbsga \leq 4.5\times 10^{-4}$. The short-dashed
line is the SM prediction, and the long-dashed and solid curve show
the ratio in the model III for $\theta=0^0, 30^0$, respectively.
The dot-dashed curve shows the same ratio at the leading order in
the model II. From the Fig.2, the lower and upper limit on $\mhp$ in
the model III can be read out:
\beq
185 GeV \leq \mhp \leq 238 GeV, \ \ for \ \ \theta=0^0,\nonumber\\
215 GeV \leq \mhp \leq 287 GeV, \ \ for \ \ \theta=30^0\label{masslimit}
\eeq
These limits are consistent with those given in Eq.(\ref{limit2}).
If we take into account the errors of theoretical predictions in
model III, the corresponding mass limit will be relaxed by about
20 GeV.

From above analysis, we get to know that for the model III the parameter
space
\beq
&& \lambda_{ij}=0, \ \ for \ \ ij\neq tt,\ \ or \ \  bb, \nonumber\\
&& |\lambda_{tt}|= 0.3,\ \ |\lambda_{bb}|=35,\ \
 \theta=(0^0 - 30^0),\ \ \mhp=(200 \pm 100 )GeV, \label{pspace}
\eeq
are allowed by the available data. For the mass $\mhp$, searches for pair
production  at LEP have excluded masses $\mhp \le 77 GeV$ \cite{gross99}.
Combining the direct and indirect limits together, we here conservatively
consider a larger range of $100$ GeV $ \leq \mhp \leq 300$ GeV, while take
$\mhp=200$ GeV as the typical value.

\section*{III.  Inclusive charmless b quark decays }

In this section, we will calculate the new physics contributions to
the two-body and three-body inclusive charmless decays of b quark
induced by the charged Higgs gluonic penguin diagrams in the models
II and III.

\subsection*{A.  $b\to s \;  gluon $ decay }

The branching ratio of  $\bsg$ at the leading order can be written as,
\beq
\brbsg =\frac{|V^*_{ts}V_{tb}|^2}{|V_{cb}|^2} \frac{8 \alpha_s(\mu)}{
\pi f(z) \kappa(z)} |C_{8G}(\mu)|^2 BR(b \to ce\bar{\nu}),
\label{brbsg1}
\eeq
with
\beq
C_{8G}(\mu)^{SM} &=& \eta^{\frac{14}{23}} C_{8G}(M_W)^{SM}
    + \sum_{i=1}^8 \bar{h}_i \eta^{a_i}, \label{c8sm}\\
C_{8G}(\mu)^{III} &=& \eta^{\frac{14}{23}} C_{8G}^R(M_W)
    + \sum_{i=1}^8 \bar{h}_i \eta^{a_i}, \label{c8r3}
\eeq
where $\eta=\alpha_s(M_W)/\alpha_s(\mu)$ with $\mu = O(m_b)$, and the
numbers $a_i$ and $\bar{h}_i$ can be found in \cite{buras97}.
The factor $\kappa(z)$ contains the QCD correction to the
semileptonic decay rate $BR(b \to c e\bar{\nu})$
\cite{cabibbo79,nir89,kim89}. To a good approximation the
$\kappa(z)$ is given by \cite{kim89}
\beq
\kappa(z) =  1- \frac{2 \alpha_s(\mu)}{3 \pi}
\left [ \left ( \pi^2 -\frac{31}{4} \right ) (1-z)^2
+ \frac{3}{2} \right ].\label{kapaz}
\eeq
And an exact analytic formula for $\kappa(z)$ can be found in
ref.\cite{nir89}.

For $b\to d g$ decay, one simply substitutes $V_{ts}^*$ by $V_{td}^*$
in Eq.(\ref{brbsg1}). For the model II, one simply replaces
$C_{8G}(\mu)$ in Eq.(\ref{brbsg1}) with $C_{8G}^{II}$ as given
in \cite{hou94}.

Fig.3 shows the branching ratios of $\brbsg$ in the SM and the models
II and III, assuming $\lambda_{tt}=0.3$, $\lambda_{bb}=35$, and
$\theta=0^0$, $30^0$. The dots line in Fig.3 is the SM prediction
$\brbsg=0.27\%$, while the short-dashed curve shows the branching
ratio $\brbsg=0.81\%$ in the model II assuming $\tan{\beta}=2$ and
$\mhp=200$ GeV. In the model III, the enhancement to the ratio
$\brbsg$ can be as large as an order of magnitude: $\brbsg \approx
2.34\%$, $4.84\%$ for $\mhp=200$, $100$ GeV respectively, as
illustrated by the long-dashed and solid curves in Fig.3. The model
III is clearly more promising than the model II to provide a large
enhancement to the decay $\bsg$. Although the current enhancement is
still smaller than $\sim 10\%$ as expected, for example in
Refs.\cite{kagan98,tseng98}, such a significant increase is obviously
very helpful for us to provide a reasonable solution for the problems
such as the `` missing charm puzzle" or the deficit $\bsl$, as being
discussed below.

\subsection*{B. Three-body charmless b quark decays}

Within the SM, the three-body inclusive charmless b quark decays have
been calculated at LO and NLO level for example in
refs.\cite{lenz97,xiao992,hou88}.  In Ref.\cite{lenz97},
Lenz et al. took into account the NLO QCD corrections from the
gluonic penguin diagrams with insertions of $Q_2$ and the diagrams
involving the interference of the $Q_{8G}$ with
$Q_{1-6}$ \cite{lenz97}.

The standard theoretical frame to calculate the decays $b \to s q \bar{q}$
for $q\in{\{u,d,s\}}$ is based on the effective Hamiltonian\cite{slac504},
\beq
{\cal H}_{eff}(|\Delta B|=1) = \frac{G_F}{\sqrt{2}} \left \{
\sum_{j=1}^2 C_j \left ( v_c Q_j^c + v_u Q_j^u \right )
- v_t \left [ \sum_{j=3}^6  C_j Q_j  + C_8 Q_{8G} \right ] \right \}
+ h.c.,
\label{heff2}
\eeq
where $v_q = V_{qs}^* V_{qb}$ and the corresponding operator basis
reads:
\beq
Q_1&=& (\bar{s}_\alpha q_{\beta})_{V-A} (\bar{q}_{\beta}b_{\alpha})_{V-A},
\label{q1}\\
Q_2&=& (\bar{s}_\alpha q_{\alpha})_{V-A} (\bar{q}_{\beta}b_{\beta})_{V-A},
\label{q2}
\eeq
with $q=u$ and  $q=c$, and
\beq
Q_3&=& (\bar{s}_\alpha b_{\alpha})_{V-A} \sum_{q=u,d,s}
(\bar{q}_{\beta}q_{\beta})_{V-A},\label{q3} \\
Q_4&=& (\bar{s}_\alpha b_{\beta})_{V-A} \sum_{q=u,d,s}
(\bar{q}_{\beta}q_{\alpha})_{V-A}, \label{q4} \\
Q_5&=& (\bar{s}_\alpha b_{\alpha})_{V-A} \sum_{q=u,d,s}
(\bar{q}_{\beta}q_{\beta})_{V+A},\label{q5} \\
Q_6&=& (\bar{s}_\alpha b_{\beta})_{V-A} \sum_{q=u,d,s}
(\bar{q}_{\beta}q_{\alpha})_{V+A}, \label{q6} \\
Q_{8G}&=& -\frac{g_s}{8\pi^2}m_b \bar{s}_\alpha \sigma^{\mu \nu}
(1+ \gamma_5)T^a_{\alpha \beta} b_{\beta} G^a_{\mu \nu}
\label{q8}
\eeq
where the $Q_1$ and $Q_2$ are current-current operators, the $Q_3 - Q_6$
are QCD penguin operators, while the $Q_{8G}$ is the chromo-magnetic dipole
operator.

For the SM part, we will use the formulae presented in \cite{lenz97}
directly. For the new physics part in the models II and III under
study here, we take into account the
new contributions from charged-Higgs gluonic penguins by using the
Wilson coefficient $C_{8G}(\mu)^{III}$ as given in Eq.(\ref{c8r3})
in the calculation, this coefficient comprises both the SM and the
new physics contributions. All other Wilson coefficients remain
unmodified.

When the NLO QCD corrections are included, one usually expand the decay
width to order $\alpha_s$,
\beq
\Gamma ( b\to s q \bar{q} ) &=& \Gamma^{(0)}+ \frac{\alpha_s (\mu)}{4 \pi}
 \left (  \Delta \overline{\Gamma}_{cc}
 + \Delta \overline{\Gamma }_{peng} + \Delta \overline{\Gamma}_W
 + \Delta \Gamma_8 \right ) + O(\alpha_s^2), \label{gammanlo}
\eeq
where $\Gamma^{(0)}$ denotes the decay rate at the LO level, while
the second part represents the NLO QCD corrections. We here use the
renormalization-scheme(RS) independent terms $\Delta
\overline{\Gamma}_{cc}$, $\Delta \overline{\Gamma }_{peng}$ and
$\Delta \overline{\Gamma}_W$.
For the convenience of the reader, the explicit expressions of
$\Delta \overline{\Gamma}_{cc}$, $\Delta \overline{\Gamma }_{peng}$ and
$\Delta \overline{\Gamma}_W$ will be given in Appendix. The
term $\Delta \Gamma_8$ in Eq.(\ref{gammanlo}) ( which will be defined
below in Eq.(\ref{gamma8}) ) is already RS independent\cite{lenz97,buras98}.
For the three-body decays $b\to d  q \bar{q}$ one simply substitutes
$s$ by $d$ in Eqs.(\ref{heff2}-\ref{gammanlo}).

At the NLO, the RS dependent Wilson coefficients $C_j(\mu)$ are given
by\cite{buras98}
\beq
C_j (\mu) = C_j^{(0)} (\mu) + \frac{\alpha_s (\mu)}{4 \pi}
              C_j (\mu)^{(1)} , \  \ j=1,\cdots,  6. \label{cjmu}
\eeq
where $C_j^{(0)}$ are the RS independent LO Wilson
coefficients, and $C_j^{(1)}$ are the RS dependent NLO corrections
\cite{buras98},
\beq
C_j^{(0)}(\mu_b) &=& \sum_{i=3}^8 k_{ji}\;  \eta^{a_i}, \label{cj0mub}\\
C_j^{(1)}(\mu_b) &=& \sum_{i=3}^8 \left
[ e_{ji}\; \eta\; E_0(x_t) + f_{ji} + g_{ji}\; \eta \right ] \eta^{a_i},
\label{cj1mub}
\eeq
where $\eta =\alpha_s(M_W)/\alpha_s(\mu_b)$, $x_t=m_t^2/M_W^2$,
the function $E_0(x_t)$ and all the numbers $a_i$, $k_{ji}$,
$e_{ji},$ $f_{ji}$, and $g_{ji}$
can be found in \cite{buras98}. The NLO QCD correction
$ C_j^{(1)}$ is RS dependent and can be split into two parts:
\beq
C_j (\mu)^{(1)} = \sum_{k=1}^6 J_{jk} C_k^{(0)} (\mu)
+ \overline{C}_j (\mu)^{(1)},\ \ \ \ j=1,\cdots , 6. \label{cjmu1}
\eeq
where parameters $J_{jk}$  are usually RS dependent,
$\overline{C}_j(\mu)^{(1)}$
is RS independent, and the precise definitions of the terms in
Eq.(\ref{cjmu1})  can be found for example in \cite{buras93}.
The terms involving $J_{jk}$ will be absorbed into $\Delta
\overline{\Gamma}_{cc}$ and $\Delta \overline{\Gamma}_{peng}$ to make the
latter scheme independent.

In the leading order the decays $b\to s s \bar{s},\;  s d \bar{d},\;
d s \bar{s}$ and $ d d\bar{d}$ are penguin-induced processes
proceeding via $Q_{3-6}$ and $Q_{8G}$, while  $b\to d u \bar{u}$
and $b\to s u \bar{u} $ also receive contributions from $Q_1$ and
$Q_2$. Combining both cases, the decay width at the LO level can be
written  as \cite{lenz97}
\beq
  \Gamma^{(0)} &=& \frac{G_F^2 m_b^5}{64 \pi^3} \left \{ \,t
  \sum_{i,j=1}^2 | v_u |^2 C_i^{(0)} C_j^{(0)} b_{ij} +
  \sum_{i,j=3}^6 | v_t |^2 C_i^{(0)} C_j^{(0)} b_{ij} \right. \nonumber\\
&& \left.
 - 2 t \sum_{ \scriptstyle i = 1,2 \atop
           \scriptstyle j = 3, \ldots 6} C_i^{(0)} C_j^{(0)} \,
  \real \left( v_u v_t^* \right) \, b_{ij} \right \}  \label{gammat}
\eeq
with $t=1$ for $q=u$ and $t=0$ for $q=d,s$. The coefficients $b_{ij}$ read
\beq
b_{ij} = \frac{16 \pi^3}{m_b^6 } \int \; d\Phi_3\; (2\pi)^4 \;
   \overline{ \langle Q_{i} \rangle^{(0)} \langle Q_{j} \rangle^{(0)\,*} }
\; =\; b_{ji} \label{bijdef}
\eeq
with $Q_{1,2}=Q_{1,2}^u$ here.  Setting the final state quark masses
to zero one finds\cite{lenz97}
\beq
b_{ij} &=& \left\{
\begin{array}{ll}
1+r/3   & \mbox{ for } i,j \leq 4,\mbox{ and } i+j \mbox{ even },  \\
1/3 +r & \mbox{ for } i,j \leq 4,\mbox{ and } i+j \mbox{ odd },
\end{array} \right. \nonumber \\
b_{55} &=& b_{66} = 1 \; ,\ \  b_{56}= b_{65}  = 1/3. \label{bij}
\eeq
Here $r=1$ for the decays $b\to d d \bar{d}$ and $b\to
s s \bar{s}$, in which the final state contains two identical
particles, and $r=0$ otherwise.  The remaining $b_{ij}$'s are zero.

Now we turn to study the contributions from the interference of
the tree diagram with $Q_8$ with operators $Q_{1-6}$, as shown in
Fig.3 of Ref.\cite{lenz97}. The tree-level correction
$\Delta \Gamma_8$ is already at the order of $\alpha_s$ and
is given by
\beq
\Delta \Gamma_8 &=& \frac{G_F^2 m_b^5}{32 \pi^3} \,  \real \!
\left[ - t \, v_u^* v_t C_{8G}(\mu)^{III}
\sum_{j=1}^2 C^{(0)}_j b_{j8} + | v_t |^2 C_{8G}(\mu)^{III}
     \sum_{j=3}^6 C^{(0)}_j b_{j8} \right]. \label{gamma8}
\eeq
in the model III, where $C_{8G}(\mu)^{III}$ has been given in
Eq.(\ref{c8r3}) with $\mu=O(m_b)$. For the case of the SM and model
II, simply replace $C_{8G}(\mu_b)^{III}$ with the appropriate
$C_{8G}(\mu_b)$. The definitions and numerical values of
coefficients $b_{j8}$ can be found in \cite{lenz97}. As mentioned
previously, the Wilson coefficient $C_{8G}^{III}$ now comprises the
contributions from both the W-penguin and the charged-Higgs penguin
diagrams. In this way, the new physics contributions are  taken into
account.

For the b quark decay rates one usually normalize them to the
semileptonic decay rate of b quark,
\beq
r_{q l} &=& \frac{\Gamma(b \to q l \bar{\nu}_l)}{\Gamma(b
\to c e \bar{\nu}_e)}, \ \ \
r_{q g} = \frac{\Gamma(b \to q\; g )}{\Gamma(b
\to c e \bar{\nu}_e)}, \nonumber\\
r_{q_1 q_2 \bar{q}_3 } &=& \frac{\Gamma(b \to q_1\; q_2 \bar{q_3} )
}{\Gamma(b \to c e \bar{\nu}_e)}, \ \ \
r_{sgg}=\frac{\Gamma(b \to s gg )}{\Gamma(b \to c e \bar{\nu}_e)},
\label{rsgg}
\eeq
for the sake of eliminating the factor of $m_b^5$ common to all b
decay rates. One also define the charmless decay rate of b quark
as
\beq
\rnc &=& \sum_{q=u,d,s}( r_{d q\bar{q}} +r_{s q\bar{q}} )
+ r_{sg} + r_{dg} +r_{sgg} + 2 r_{ue} + r_{u\tau} \label{rnc}
\eeq
where rare radiative decays, for example  $b \to s \gamma $, have
been neglected. To order $\alpha_s$, the semileptonic decay rate
takes the form
\beq
\Gamma(b \to c e\bar{\nu}_e) = \frac{G_F^2 m_b^5}{192 \pi^3}
|V_{cb}|^2 f(z) \kappa(z)
\eeq
where the factors $f(z)$ and $\kappa(z)$ have been given in
Eqs.(\ref{fz}) and (\ref{kapaz}).

To calculate $\rnc$ we also need explicit expressions of $r_{ue}$,
$r_{sg}$, $r_{dg}$ and $r_{sgg}$. For $r_{ue}$ one finds \cite{bigi92},
\beq
r_{ue}= \frac{|V_{ub}|^2}{|V_{cb}|^2} \frac{1}{f(z)}
\left \{ 1 + \kappa(z) - \kappa(0)
+ 6 \left [ \frac{(1-z^2)^4}{f(z)} -1 \right ]
\frac{\lambda_2}{m_b^2} \right \}, \label{eq:rue}
\eeq
where $\lambda_2=0.12 GeV^2$ encodes the chromomagnetic interaction
of the b quark with light degrees of freedom, and the factors of
$f(z)$ and $\kappa(z)$ have been given in Eqs.(\ref{fz}) and
(\ref{kapaz}).

From Eq.(\ref{brbsg1}), we  get
\beq
r_{sg}&=& \frac{|V^*_{ts}V_{tb}|^2}{|V_{cb}|^2} \frac{8 \alpha_s(\mu)}{
\pi f(z) \kappa(z)}|C_{8G}(\mu)|^2,\\ \label{eq:rsg}
r_{dg}&=& \frac{|V^*_{td}V_{tb}|^2}{|V_{cb}|^2} \frac{8 \alpha_s(\mu)}{
\pi f(z) \kappa(z)}|C_{8G}(\mu)|^2. \label{eq:rdg}
\eeq

For $r_{sgg}$, we use the formulae as given in \cite{hou88,xiao992},
\beq
r_{sgg}= \frac{1}{|V_{cb}|^2}\frac{3 \alpha_s(\mu)^2}{16 \pi^2
f(z)\kappa(z)} |\sum_{i=u,c,t}V_{is}^*V_{ib}f_1(x_i,q^2)|^2, \label{eq:rsgg}
\eeq
where $x_i=m_i^2/M_W^2$, the functions $f_1(x_i,q^2)$ can be found
for example in \cite{xiao992}. In the numerical
calculation, we assume that $q^2=m_b^2/2$. Since the new
contribution to the decay $b \to s g g$ due to the charged Higgs
penguin is negligibly small \cite{xiao992}, we do not consider
the new physics corrections to this decay here.
In Ref.\cite{lenz97}, the authors did not include $r_{sgg}$ in the
estimation of $\rnc$. We here will include this mode, since its
branching ratio is rather large \cite{hou88,xiao992},
as shown in the  \tab{tab:theta0}.

The corresponding branching ratios for two-body and three-body
charmless b decays are defined as
\beq
BR(b \to X ) &=& r_{X}\, \cdot \, BR(b \to c\, e\bar{\nu}_e )^{exp},
 \label{eq:brb2}
\eeq
where ratios $r_{X}$  have been defined previously. In the numerical
calculations, $BR( b \to c\, \bar{\nu}_e)^{exp}=10.70\%$ will be used
\footnote{For more details, one can see the discussions about the
semileptonic branching ratios of b decay in next section.}.

By using the input parameters as given in Eq.(\ref{sip}) and assuming
$|\lambda_{tt}|=0.3$,  $|\lambda_{bb}|=35$, $\mhp=200$GeV and
$\theta=0^0$ or $30^0$, we find the numerical results of the
decay rates  and  the branching ratios  for  various charmless b
quark decays and collect them  in \tab{tab:theta0}. We also show
the corresponding results in the  model  II assuming $\mhp =200$ GeV
and $\tan{\beta}=2$. For larger $\tan{\beta}$ the new physics
contributions in model II will  become smaller. $\Delta BR$ in
\tab{tab:theta0} is defined as
\beq
\Delta BR(b \to X) = \left [ BR(b\to X) - BR(b \to X)^{SM} \right ]/
BR(b \to SM)^{SM} \label{dbr}
\eeq

\begin{table}[tbp]
\caption{The rates $r$ and branching ratios in the SM and models II
and III, assuming $|\lambda_{tt}|=0.3$,  $|\lambda_{bb}|=35$,
$\mhp=200$GeV, $\tan{\beta}=2$,  and  $\theta=0^0$ or $30^0$(the
numbers in parenthesis). We also use
$BR(B \to X_c e \bar{\nu}_e)^{exp}  =10.70\%.$ as given in
Eq.(\ref{bslexp}). }
\begin{center}
\begin{tabular}{l cc cccc} \hline \hline
& \multicolumn{2}{c}{SM} &
  \multicolumn{3}{c}{Model\ \ \  III}& Model\ \ II  \\ \cline{2-7}
decay mode  & $ r $ & $BR (\%)$ &
$r $ &$BR (\%)$ & $\Delta BR (\%)$& $\Delta BR (\%)$ \\ \hline
 $b \to d\,u\bar{u}$&0.051 & 0.545  &0.052   & 0.554&1.6 &-0.3 \\
&&& (0.053)&(0.571)&(4.7)& \\
 $b \to d\,d\bar{d}$&0.0005 & 0.006&0.00078 & 0.103&68.2&-13.4  \\
&&& (0.0007)&(0.010)&(59.0)& \\
 $b \to d\,s\bar{s}$&0.0006 & 0.005&0.00096 & 0.008&68.7&-13.5  \\
&&& (0.0009)&(0.008)&(59.5)& \\ \hline
 $b \to s\,u\bar{u}$&0.018 & 0.192 &0.027 & 0.286&49.0&-9.6  \\
&&& (0.0237)&(0.255)&(32.8)& \\
 $b \to s\,d\bar{d}$&0.019 & 0.206 &0.030 & 0.322&56.0&-11.0  \\
&&& (0.0285)&(0.307)&(48.5)& \\
 $b \to s\,s\bar{s}$&0.016 & 0.168 &0.024 & 0.262&56.7&-9.9  \\
&&& (0.0232)&(0.250)&(49.2)& \\ \hline
 $b \to s\,g$&0.025   & 0.270      &0.192 & 2.065&663.6&202.3  \\
&&& (0.217)&(2.339)&(765.0)& \\
 $b \to d\,g$&0.00092 & 0.010      &0.007 & 0.070&663.6&202.3  \\
&&& (0.008)&(0.086)&(765.0)& \\
 $b \to s\,g g$&0.070 & 0.757      &0.070 & 0.757&-&- \\ \hline
 $b \to u\,e \bar{\nu}_e$&0.013 & 0.144   &0.013 & 0.144&-&- \\
 $b \to u\,\mu \bar{\nu}_{\mu}$&0.013 & 0.144   &0.013 & 0.144&-&- \\
 $b \to u\,\tau \bar{\nu}_{\tau}$&0.004 & 0.0004   &0.004 & 0.0004&-&-
\\ \hline
 $ b \to no\ \ charm$ & 0.23&2.49 &0.43 & 4.67&87.6&20.3  \\
                      &     &     &(0.46) & (4.91)&(97.3)&      \\
\hline \hline
\end{tabular} \end{center} \label{tab:theta0} \end{table}

Fig.\ref{fig:fig4} shows the mass dependence of the branching ratios
$BR(b \to s\, q \bar{q})$ with $q\in \{u,d,s\}$ in the SM and
model III,  using the input parameters in Eq.(\ref{sip}) and assuming
$|\lambda_{tt}|=0.3$,  $|\lambda_{bb}|=35$, and
$\theta=30^0$. In Fig.\ref{fig:fig4}, the three curves ( horizontal
lines) are the theoretical predictions in the model III ( SM ) for
$q=u,d,s$, respectively. For $\mhp =200GeV$, as listed in
\tab{tab:theta0}, the enhancement
to the decay mode $b \to d \, u \bar{u}$ is only $4.7\%$, but  the
enhancements to other five three-body b quark decay modes are rather
large: from $\sim 30\%$ to $\sim 70\%$. In the model II, however,
the new contributions are negative and will decrease the branching
ratios slightly, from $-0.3\%$ to $-13.5\%$ for different decay
modes.

Fig.\ref{fig:fig5} shows the branching ratio $BR(b \to no \ \
charm) $ in the SM and models II and III, using the input parameters
in Eq.(\ref{sip}) and assuming $|\lambda_{tt}|=0.3$,  $|\lambda_{bb}|=35$,
and $\theta=0^0$, $30^0$. The dots line in Fig.\ref{fig:fig5} is
the SM prediction $\brnoc=2.49\%$. The short-dashed curve shows the
the ratio in the model II, $\brnoc =2.98\%$ ($3.23\%$) for $\mhp=200$
(100) GeV and $\tan{\beta}=2$. The long-dashed and solid curve
show the theoretical predictions in the model III: $\brnoc =4.67\%$
($4.91\%$) for $\mhp=200$ GeV and $\theta=0^0, 30^0$, respectively.
For the model III with $\mhp=100$ GeV, one finds that
$\brnoc =7.27\%$ ($7.60\%$) for $\theta=0^0, 30^0$, respectively.

It is easy to see from Fig.\ref{fig:fig5} and \tab{tab:theta0} that
the new physics enhancement to the branching ratios of three-body
charmless b quark decays in the model III is much larger than that
in model II within the parameter space considered.

\section*{ IV.  $\nc$ and $\bsl$ }

The ratio $\bsl$ is the average over weakly-decaying hadrons
containing one b quark. For the CLEO experiments running on the
$\Upsilon (4S)$ resonance, the average is over $B^+$ and $B^0$ and
their charge conjugate hadrons. For the experiments running on
$Z^0$ resonance, however, the average is over $B^+$, $B^0$, $B^0_s$
and $N_b$\footnote{$N_b$ is in turn the mixture of $\Lambda_b (udb)$,
$\Sigma_b (usb)$, $\Xi_b(dsb)$ and $\Omega_b (ssb)$.}.

The charm multiplicity $\nc$ is the average over the b-hadrons
produced in the given environment. CLEO and LEP collaborations
presented new measurements of inclusive $b \to c $ transitions that
can be used to extract $\nc$. One naively expect $\nc=1.15$ with
the additional $15\%$ coming from the tree-level decay chain
$b \to u W^- \to u\, \bar{c} s $. This expectation can be verified
experimentally by adding all inclusive $b \to c$ branching ratios,
and counting twice for the decay modes with 2 charm quarks  in the
final state.

In this section, we will investigate the new physics contributions,
induced by the charged Higgs penguins in the models II and III, to
the ratio $\bsl$ and the charm multiplicity $\nc$.

\subsection*{A. $\nc$ and $\bsl$: experimental measurements }

The $\bsl$ deficit was first point out in around 1994 \cite{bigi94}
when the theoretical prediction was considered to be difficult
to produce $\bsl \leq 12\%$ while the 1995 CLEO data on
$\Upsilon (4S)$ resonance was $\bsl =(10.49 \pm 0.46) \%$ \cite{cleo96}.
In the following, we use the 1998 Particle Data Group
value \cite{pdg98}
\beq
\bsl =(10.45 \pm 0.21) \% \label{cleobsl}
\eeq
as the measured $\bsl$ on $\Upsilon (4S)$.

For the experiments on the $Z^0$-peak, all the four LEP
collaborations \cite{aleph95,delphi99,l399,opal99} reported their
measured values of the ratio $\bsl$ as listed in \tab{tab:bslexp}.
The seventh row shows the averaged result of the ratio $\bsl$ on
$Z^0$-peak \footnote{ We here made an arithmetic average
over four results as done in \cite{yamamoto99}, but
the newest L3 data \cite{l399} has been used here in the average. }:
$\bsl^b = (10.66 \pm 0.17)\%$. This $\bsl^b$ on the
$Z^0$-peak can be converted to $\Upsilon (4S)$ value by multiplying
a factor of $\tau_B/\tau_b= 1.026$: $\bsl = (10.94 \pm 0.19)\%$
($Z^0$  corrected). In fact, there is still a $2\,\sigma$ discrepancy
in ratio $\bsl$ between the high energy $Z^0$ value and the low
energy $\Upsilon (4S)$ value. The average of the $Z^0$ and
$\Upsilon (4S)$ values of $\bsl$ is
\beq
\bsl = (10.70 \pm 0.21)\%\ \ {\rm Overall \ \ average}\label{bslexp}
\eeq
where we conservatively chose $0.21$ as the overall error of
the measured $\bsl$.

\begin{table}[tbp]
\caption{Recent CLEO and LEP measurements of the ratio $\bsl$. }
\begin{center}
\begin{tabular}{ll} \hline \hline
$\bsl\; (\%)$ & Experiment \\ \hline
$10.45 \pm 0.21 $ & $ \Upsilon (4S) $ PDG98 \cite{pdg98} \\  \hline
$11.01 \pm 0.10 (stat.) \pm 0.30(syst.)$ &  ALEPH 95 \cite{aleph95} \\
$10.65 \pm 0.07 (stat.) \pm 0.25(syst.) ^{+0.28}_{-0.12} (model) $
&  DELPHI 99 \cite{delphi99} \\
$10.16 \pm 0.13(stat.) \pm 0.30(syst.) $ &  L3  99 \cite{l399} \\
$10.83 \pm 0.10(stat.) \pm 0.20(syst.) ^{+0.20}_{-0.13}(model)  $
&  OPAL 99 \cite{opal99} \\  \hline
$10.66 \pm 0.17 $ &  $Z^0$-{\rm peak} \\
$10.94 \pm 0.19 $ &  $Z^0$  {\rm corrected} \\
$10.70 \pm 0.21 $ & {\rm overall\ \  average} \\  \hline \hline
\end{tabular}
\end{center} \label{tab:bslexp}
\end{table}

As for the charm counting, the value of $\nc$ measured at the
$\Upsilon(4S)$ \cite{cleo97} is still smaller than that measured at
$Z^0$-peak \cite{pdg98}:
\beq
\nc &=& \left \{ \begin{array}{ll}
1.10 \pm 0.05, &  \Upsilon(4S), \\
1.20 \pm 0.07, & Z^0{\rm -peak}
\\ \end{array} \right.\label{ncexp1}
\eeq
The average of the $\Upsilon (4S)$ and $Z^0$ result leads to
\beq
\nc &=&1.14 \pm 0.04, (Z^0 + \Upsilon (4S) ) \label{avncexp}
\eeq

\subsection*{B.  $\nc$ and $\bsl$: theoretical predictions}

Within the SM, the basis of the prediction for $\bsl$ and $\nc$
is the assumption of quark-hadron duality. The estimation
for various inclusive decay rates is usually performed by using
the heavy-quark expansion(HQE) \cite{neubert94} and the perturbative
QCD in the framework of operator product expansion. The HQE allows to relate
the inclusive decay rate of  B meson to that of the underlying $b$
quark decay process: $ \Gamma(B \to X) = \Gamma(b \to x)
+ O(1/m_b^2)$.

The theoretical prediction for $\bsl$ with the inclusion of the
$O(\alpha_s)$ QCD corrections and the hadronic corrections to the
free quark decay of order $1/m_b^2$ is currently available
\cite{bslsm,neubert97}.
The $\bsl$ and $\nc$ can be defined as \cite{bslsm,neubert97}
\beq
\bsl &=& \frac{1}{ \sum_l \, r_{cl} + r_{c\bar{u}d} + r_{c\bar{c}s}
+ \rnc }, \label{bsl1} \\
\nc &=& 1+ \frac{ r_{c\bar{c}s} - \rnc}{\sum_l \, r_{cl}
+  r_{c\bar{u}d} +  r_{c\bar{c}s} + \rnc }, \label{nc1}
\eeq
where $r_{ce}=r_{c\mu}=1$, $r_{c\tau}=0.25$, and $r_{c\bar{u}d}$
($r_{c\bar{c}s}$) is the rate of the decay mode $b \to c\bar{u}d'$
($b \to c \bar{c}s'$) where $d'$ ($s'$) is the appropriate Cabibbo
mixture of $d$ and $s$ quarks.

The $\rnc$ has been defined and calculated in last section.
In the SM, we have
\beq
\rnc= 0.23 \pm 0.08, \label{rncsm}
\eeq
where the error mainly comes from the uncertainties of the scale
$\mu$ and the mass ratio $m_c/m_b$ \cite{lenz97}.

As is well known, the main difficulty in calculating $\bsl$ and
$\nc$ is in the non-leptonic branching ratios $r_{c\bar{u}d}$ and
$r_{c\bar{c}s}$. For $r_{c\bar{u}d}$, a complete NLO calculation
has been performed \cite{bslsm} which gives
\beq
r_{c\bar{u}d}=4.0 \pm 0.4, \label{rcud}
\eeq
where the error mainly comes from the uncertainties of the scale
$\mu$, the quark masses $m_c$ and $m_b$ and the assumption of
quark-hadron duality \cite{bslsm}. Furthermore, the error of
the estimation for $r_{c\bar{c}s}$ is generally considered to be
larger than that for $r_{c\bar{u}d}$. The
enhancement of $b \to c\bar{c}s$ due to large QCD corrections is
about $30\%$ \cite{bslsm}. Such enhancement will decrease the value
of $\bsl$, but increase the size of $\nc$.

Using the on-mass-shell sheme, the SM theoretical predictions for
$\bsl$  and $\nc$ at the NLO level are
\beq
\bsl &=& (12.0 \pm 0.7 \pm 0.5 \pm 0.2 ^{+0.9}_{-1.2})\%, \label{bslth1}\\
\nc &=& 1.24 \mp 0.05\pm 0.01, \label{ncth1}
\eeq
as given in Ref.\cite{bslsm}\footnote{The last and largest error
of $\bsl$ comes from the uncertainty of the renormalization scale
$\mu$; while the main error of $\nc$ is the the uncertainty in
$m_b$ \cite{bslsm}.}; and
\beq
\bsl &=& \left \{ \begin{array}{ll}
(12.0 \pm 1.0) \% &  \mu =m_b, \\
(10.9 \pm 1.0) \% & \mu =m_b/2, \\ \end{array} \right.\label{bslth2} \\
\nc &=& \left \{ \begin{array}{ll}
1.20 \mp 0.06  &  \mu =m_b, \\
1.21 \mp 0.06  &  \mu =m_b/2, \\ \end{array} \right.\label{ncth2}
\eeq
as given in Ref.\cite{neubert97} with the error mainly result from
the variation of the scale $\mu$ and $m_c/m_b$.

Comparing the observed and predicted values of $\bsl$ and $\nc$,
one can see that: (a) after considering all the corrections, the
theoretical values of $\bsl$ now come down and more or less
consistent with the measurement, but unfortunately at the expense
of boosting $\nc$;
(b) the central value of $\nc$ in Ref.\cite{bslsm} is higher than
that in Ref.\cite{neubert97}, although two predictions are agree
within errors;
(c) there is still $2.8\,\sigma$ discrepancy between the $\nc$
measured by CLEO and the theoretical prediction \cite{bslsm}:
$1.10 \pm 0.05$ against $1.24 \pm 0.05$

If we'd like to drop down the large uncertainty in the
calculation for $b\to c \bar{c}s'$ decay mode, we can eliminate
the ratio $r_{c\bar{c}s'}$ from the expression of $\bsl$ and $\nc$
and find,
\beq
\nc= 2 - \left ( 2.25+ r_{c \bar{u}d} + 2 \rnc \right ) \bsl
\label{ncthb}
\eeq
which is a linear correlation between $\bsl$ and $\nc$. Using the
values for $\bsl$ (\ref{bslexp}), $r_{c\bar{u}d}$ (\ref{rcud}),
and $\rnc$ (\ref{rncsm}), one finds
\beq
\nc = 1.28 \pm 0.05, \label{ncthc}
\eeq
for $\bsl=(10.70 \pm 0.21)\%$. The overall uncertainty of this prediction of $\nc$ should be smaller
than that as given in Eqs.(\ref{ncth1}) and (\ref{ncth2}).
The $2.6\, \sigma$ discrepancy between the $\nc$ in Eq.(\ref{ncthc})
and $\nc$  measured at $\Upsilon(4S)$ motivated proposals of new
physics which will enhance $\rnc$ and in turn decrease $\nc$. That
is what we try to do here.

As shown in \tab{tab:theta0}, the ratio $\rnc$ will be increased
significantly after taking the new physics effects into account,
which will in turn decrease both $\bsl$ and $\nc$ accordingly.
From Eq.(\ref{ncthb}) and using the values for $\bsl$ (\ref{bslexp}),
$r_{c\bar{u}d}$ (\ref{rcud}), and $\rnc$ (\ref{rnc}),
one finds
\beq
\nc = \left \{ \begin{array}{ll}
1.23 \pm 0.05  &  {\rm for \ \ \mhp=200 GeV} \\
1.18 \pm 0.05  &  {\rm for \ \ \mhp=100 GeV} \\
\end{array} \right.\label{ncthnp}
\eeq
for $\theta=30^0$ and $\mu=m_b$. The $\mu$- and $\theta$-dependence
of $\nc$ is rather weak: the central value of $\nc$
will go down (up) by only $\sim 0.01$ for $\mu=m_b/2$ ($\theta=0^0$).
For $\bsl$ in the model III, the agreement between the prediction
and the data will be improved slightly by a decrease $ 0.003$
$(0.005)$ for $\mhp=200$ ($100$) GeV due to the inclusion of new physics
contributions. In the model II,
the resulted decrease for $\nc$ ($\bsl$) is only
$0.01$ ($0.001$) and plays no real role. Most importantly, one can
see from Eqs.(\ref{ncexp1},\ref{avncexp},\ref{ncthc},\ref{ncthnp})
that the  predicted $\nc$ and the measured
$\nc$ now agree within roughly one standard deviation after taking
into account the effects of gluonic charged Higgs penguins in the
model III with a relatively light charged Higgs
boson, as illustrated in Fig.\ref{fig:fig6}.

\section*{V.  Summary and discussions}

In the framework of the general two-Higgs doublet models, we
calculated the charged-Higgs penguin contributions to
(a) the rare radiative decay $b \to s \gamma$;
(b) the inclusive charmless decays $b \to q' g$ and $b \to q'\,
q \bar{q}$ with $q'\in \{d,s\}$ and $q \in\{ u,d,s\}$;
(c) the charm multiplicity $\nc$ and semileptonic branching ratio
$\bsl$.

In section II, we studied the experimental constraint on the model
III from the CLEO data of $\bsga$ decay. With the help of previous
works \cite{atwood97,atwood96,aliev99,chao99}, we found the parameter
space of the model III allowed by the available data, as shown in
Eq.(\ref{pspace}).

In section III, we firstly calculated the new physics contributions to the
decay $b \to s g$ and found that the branching ratio $\brbsg$ can
be greatly enhanced from the SM prediction of $0.27\%$ to
$ 2.34\%$ ($4.84\%$) in the model III for $\mhp=200$ ($100$) GeV,
as illustrated in Fig.3. Such a significant enhancement is clearly
very helpful to resolve the missing charm/$\bsl$ problem appeared
in B experiments.

Following the method of Ref.\cite{lenz97}, we then calculated the new
physics contributions to three-body inclusive charmless decays
of b quark due to the interference between the operators $Q_{1-6}$
and $Q_{8G}$. The Wilson coefficient $C_{8G}^{III}$ in Eq.(\ref{c8r3})
now describe the contributions from both the $W^\pm$ and $H^{\pm}$
QCD penguins, the latter is the new physics part we focus in here.
From numerical calculations. we found that: (a) the new physics
enhancement to the decay $b \to d u \bar{u}$ is only $\sim 1.6\%$
since this mode is dominated by the tree diagrams; (b) the branching
ratios of other five three-body b decay modes  are strongly enhanced
by the new charged Higgs penguins: $ 30\%$ to $ 70\%$
increase can be achieved within the considered parameter space.
The new contributions to the corresponding  branching ratios in the
model II is, however, small in size and negative in sign against
the theoretical predictions in the SM. As shown in \tab{tab:theta0}
and Fig.\ref{fig:fig5}, the ratio $\brnoc $ can be increased from
the SM prediction $\brnoc=2.49\%$ to $\brnoc=4.91\%$ ($7.60\%$)
in the model III for $\mhp=200$ ($100$) GeV.

In section IV, we studied the current status about the theoretical
predictions and experimental measurements for the semileptonic
branching ratio of B meson decay $\bsl$ and the charm multiplicity
$\nc$, and calculated the new physics contributions, induced by
the charged Higgs penguins in the model III (II), to both  $\bsl$
and $\nc$. With an enhanced ratio $\brnoc$, both the $\bsl$
and $\nc$ will be decreased accordingly:
(a) the central value of $\bsl$ can be decreased slightly by
$0.003$ ($0.005$) for $\mhp=200$ ($100$) GeV;
(b) the value of $\nc$ can be lowered significantly from the
prediction $ \nc=1.28 \pm 0.05$ in the SM to $ \nc= 1.23 \pm 0.05$,
$1.18 \pm 0.05$ for $\mhp=200, 100$ GeV, respectively.

In short, the predicted $\nc$ and the measured
$n_c$ now agree within roughly one standard deviation after taking
into account the effects of gluonic charged Higgs penguins in the
model III with a relatively light charged Higgs boson, while
the agreement between the theoretical prediction and the data for
$\bsl$ can also be improved by inclusion of new physics effects.

\section*{ACKNOWLEDGMENTS}

C.S. Li and K.T. Chao  acknowledge the support by the National Natural
Science Foundation of China,  the State Commission of Science
and technology of China and the Doctoral Program Foundation of Institution
of Higher Education. Z.J. Xiao acknowledges the support by the National
Natural Science Foundation of China under the Grant No.19575015 and
the Outstanding Young Teacher Foundation of the Education Ministry
of China.

\section*{Appendix:    RS independent  $\Delta \overline{\Gamma}_{cc}$,
$\Delta \overline{\Gamma }_{peng}$ and $\Delta
\overline{\Gamma}_W$ }

For the convenience of the reader, we here present the explicit
expressions of the RS independent NLO corrections $\Delta
\overline{\Gamma}_{cc}$, $\Delta \overline{\Gamma }_{peng}$ and
$\Delta \overline{\Gamma}_W$. For more details one can see the
original paper \cite{lenz97}.

The term $\Delta \overline{\Gamma}_{cc}$ in Eq.(\ref{gammanlo})
describes  the current-current type corrections proportional to
$C_{1,2}^{(0)} \cdot C_{1,2}^{(0)}$\cite{lenz97}
\beq
\Delta \overline{\Gamma}_{cc} &=&  t \,  \frac{G_F^2 m_b^5}{32 \pi^3}
   \, | v_u |^2 \sum_{i,j=1}^2
   C_i^{(0)} C_j^{(0)} \left [ h_{ij} + \sum_{k=1}^2 J_{ki} b_{kj} \right ]
\label{gammacc}
\eeq
with $t=1$ for $q=u$ and $t=0$ for $q=d,s$, and the coefficients
$h_{ij}$ and $J_{ki}$ can be found in \cite{lenz97}.

The term $\Delta \overline{\Gamma}_{peng}$ in Eq.(\ref{gammanlo})
describes  the effect of penguin diagrams involving $Q_{1,2}$
\cite{lenz97},
\beq
\Delta \overline{\Gamma}_{peng} &=&   \frac{G_F^2 m_b^5}{32 \pi^3}
   \, \real \left[ \, t
    \sum_{i,j=1,2} C_i^{(0)} C_j^{(0)} \,    v_u
    \left[ v_c^* g_{ij} (x_c)  + v_u^* g_{ij} (0) \right ]
     \right.  \nonumber\\
&& \left. - \sum_{ \scriptstyle i = 1,2 \atop \scriptstyle
          j = 3, \ldots 6} \! C_i^{(0)}  C_j^{(0)}
   \; v_t
    \left[ v_c^* g_{ij}(x_c)  + v_u^* g_{ij} (0) \right ] \right ]
    \nonumber\\
&& + \;  \real    \left[ - \, t \,   v_u v_t^*
    \sum_{ \scriptstyle i,j=1,2 \atop \scriptstyle
            k=3,\dots 6} C_i^{(0)} C_j^{(0)} \,
     J_{ki} b_{jk} +    | \xi_t |^2 \!
  \sum_{ \scriptstyle i = 1,2 \atop \scriptstyle
         j,k  = 3, \ldots 6} \! C_i^{(0)}  C_j^{(0)}
    J_{ki} b_{jk}   \right ]. \label{gammapeng}
\eeq
with $t=1$ for $q=u$ and $t=0$ for $q=d,s$. The explicit expressions
of coefficients $g_{ij}$ and $J_{ki}$ can
be found in \cite{lenz97}.

Finally,  $\Delta \overline{\Gamma}_{W}$ is given by
\beq
\Delta \overline{\Gamma}_{W}  &=&\frac{G_F^2 m_b^5}{32 \pi^3}  \left[
 t \sum_{i,j=1}^2 | v_u |^2  \left[ C_i^{(0)}\;
 \overline{C}_j^{(1)}  \right] b_{ij}
+ \sum_{i,j=3}^6 | v_t |^2  \left[ C_i^{(0)}\;
 \overline{C}_j^{(1)}  \right] b_{ij} \right. \nonumber\\
&& \left. - \, t \sum_{ \scriptstyle i = 1,2 \atop \scriptstyle
j = 3, \ldots, 6} \left[ C_i^{(0)}\; \overline{C}_j^{(1)}
+\; \overline{C}_i^{(1)}\; C_j^{(0)}  \right]
 Re \left( v_u^* v_t \right) \, b_{ij}  \right] \label{gammaw}.
\eeq
where $t=1$ for $q=u$ and $t=0$ for $q=d,s$, the $b_{ij}$ have been given
in Eq.(\ref{bij}).

\newpage




\newpage

\begin{figure}[t] 
\begin{minipage}[t]{0.95\textwidth}
\centerline{\epsfxsize=\textwidth \epsffile{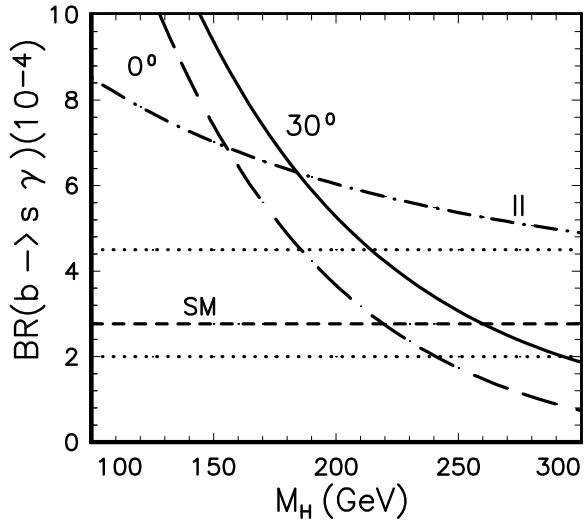}}
\vspace{-20pt}
\caption{Plots of the branching ratio $\brbsga$ versus
$\mhp$ in the SM and models II and  III. The short-dashed line is
the SM prediction, and the band between two dots lines refers to
the CLEO data.  The dot-dashed curve shows the ratio in the model
II, while the long-dashed and solid curve show the ratios in the
model III for $\theta=0^0, 30^0$, respectively.} \label{fig:fig2}
\end{minipage}
\end{figure}

\newpage
\begin{figure}[t] 
\vspace{-100pt}
\begin{minipage}[t]{0.95\textwidth}
\centerline{\epsfxsize=\textwidth \epsffile{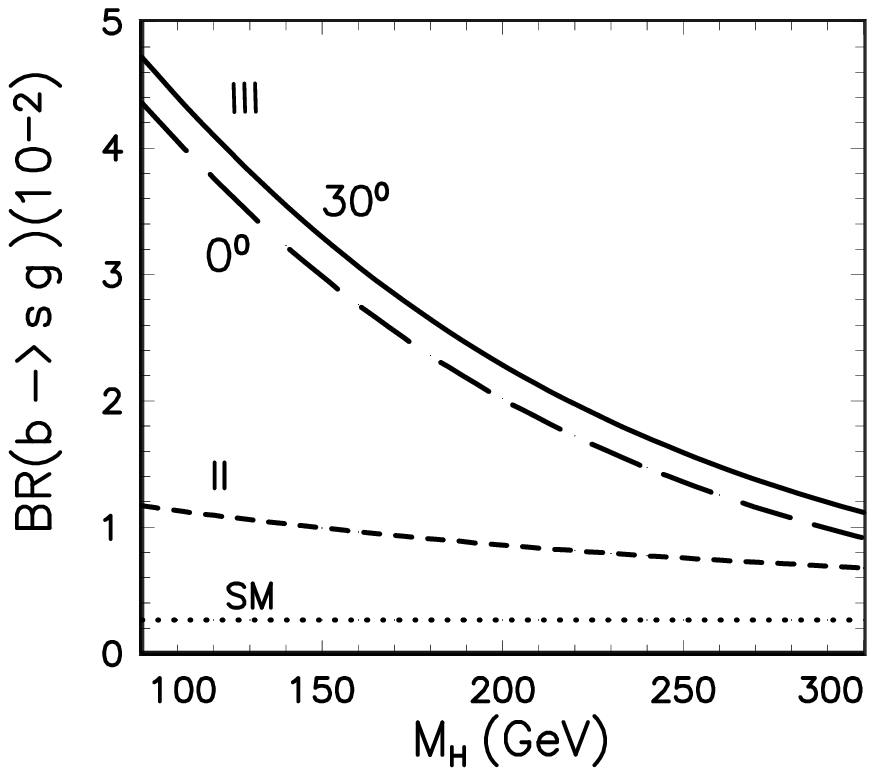}}
\vspace{-15pt}
\caption{Plots of the branching ratio $BR(b \to s g)$ versus
$\mhp$ in the SM and models II and  III. The dots line is the SM
prediction, the  short-dashed curve shows the
the ratio in the model II, and the long-dashed and solid curve
show the ratios in the  model III for $\theta=0^0, 30^0$,
respectively.}
\label{fig:fig3}
\end{minipage}
\end{figure}

\begin{figure}[t] 
\vspace{-50pt}
\begin{minipage}[t]{0.98\textwidth}
\centerline{\epsfxsize=\textwidth \epsffile{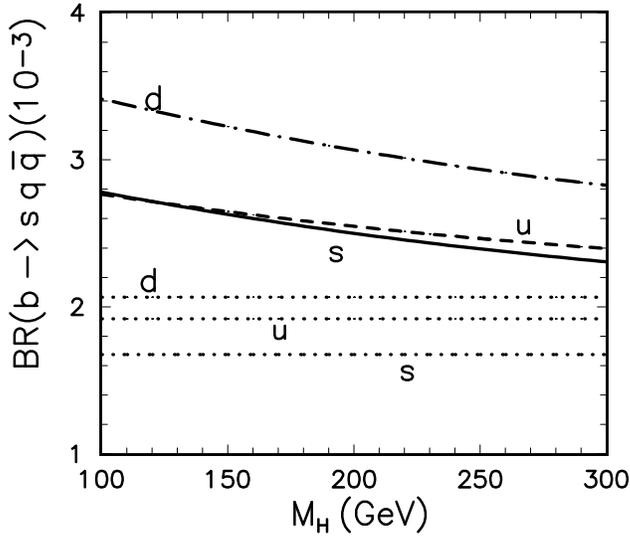}}
\vspace{-15pt}
\caption{Plots of branching ratio $BR(b \to s q \bar{q})$ versus
$\mhp$ in model III. The three curves ( horizontal lines) are the
theoretical predictions in the model III (SM) for $q=u,d,s$,
respectively.}
\label{fig:fig4}
\end{minipage}
\end{figure}

\newpage
\begin{figure}[t] 
\vspace{-100pt}
\begin{minipage}[t]{0.95\textwidth}
\centerline{\epsfxsize=\textwidth \epsffile{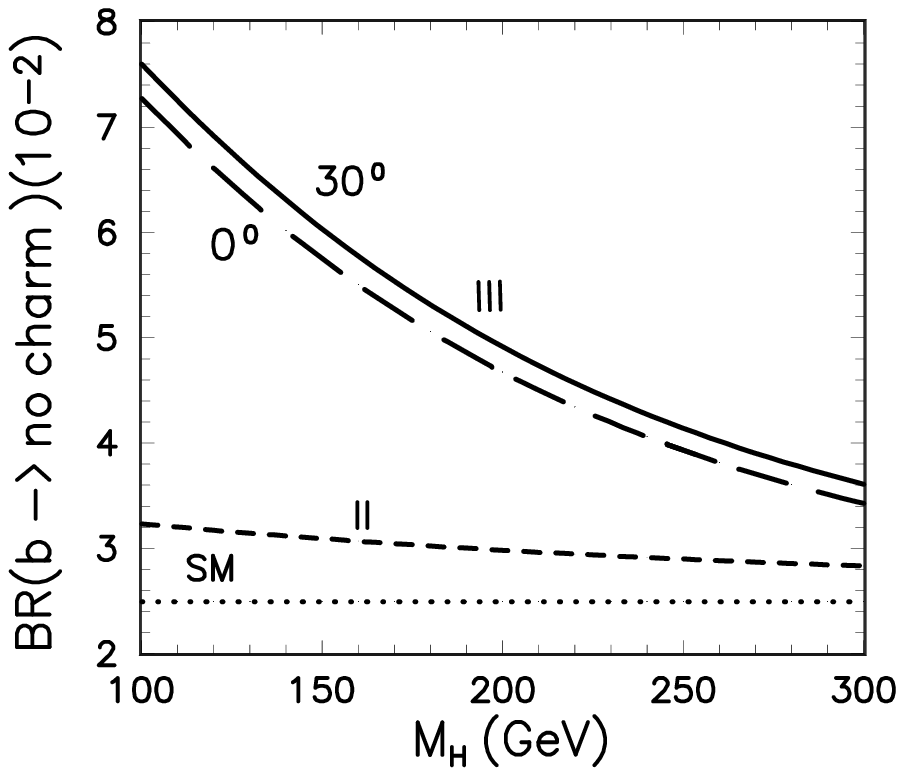}}
\vspace{-15pt}
\caption{Plots of the branching ratios $BR(b \to no \ \  charm )$
versus $\mhp$ in the SM and models II and  III. The dots line is
the SM prediction, the short-dashed curve shows the
the ratio in the model II, and the long-dashed and solid curve
show the theoretical predictions in the model III for $\theta=0^0,
30^0$, respectively. }
\label{fig:fig5}
\end{minipage}
\end{figure}

\newpage
\begin{figure}[tb] 
\vspace{-100pt}
\begin{minipage}[t]{0.95\textwidth}
\centerline{\epsfxsize=\textwidth \epsffile{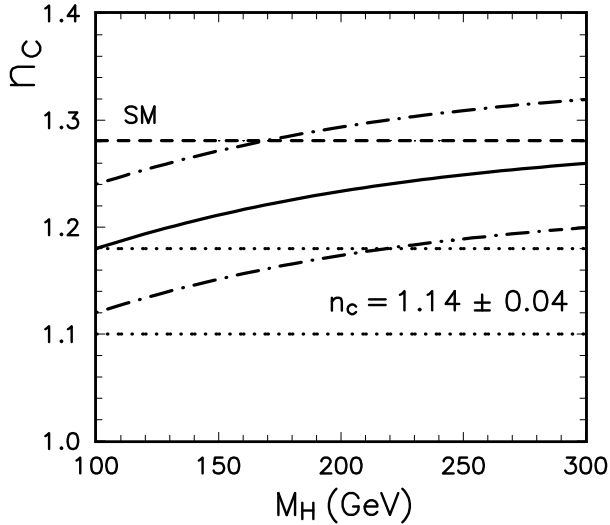}}
\vspace{-15pt}
\caption{Plots of Charm multiplicity $\nc$ versus $\mhp$ in the SM
and model III for $\bsl=10.70\%$. The short-dashed line is
the SM prediction, and the band refers to the data of $\nc
=1.14 \pm 0.04$. The solid curve, the upper and lower dot-dashed
curves together show the central value and the $1\,\sigma$ error of
the theoretical prediction for $\nc$ in the model III.}
\label{fig:fig6}
\end{minipage}\hspace{2ex}
\end{figure}

\end{document}